\documentclass[aps,pra,reprint,groupedaddress,floatfix]{revtex4-1} 
\pdfoutput=1
\usepackage{graphicx}
\usepackage{float}
\newcommand{\Yb}{$^{171}$Yb$^+$}

\begin{document}

\title{Monolithic Microfabricated Symmetric Ion Trap for Quantum Information Processing}

\author{Fayaz Shaikh, Arkadas Ozakin, Jason M. Amini, Harley Hayden, C.-S. Pai, Curtis Volin, Douglas R. Denison, Daniel Faircloth, Alexa W. Harter, and Richart E. Slusher}

\affiliation{Quantum Information Systems Group\\Georgia Tech Research Institute}

\date{May 24, 2011}

\begin{abstract}
We describe a novel monolithic ion trap that combines the flexibility and scalability of silicon microfabrication technologies with the superior trapping characteristics of traditional four-rod Paul traps. The performace of the proposed microfabricated trap approaches that of the macroscopic structures. The fabrication process creates an angled through-chip slot which allows backside ion loading and through-laser access while avoiding surface light scattering and dielectric charging.  The trap geometry and dimensions are optimized for confining long ion chains with equal ion spacing [G.-D. Lin, {\em et al.}, Europhys. Lett. {\bf 86}, 60004 (2009)].  Control potentials have been derived to produce linear, equally spaced ion chains of up to 50 ions spaced at 10~$\mu$m.  With the deep trapping depths achievable in this design, we expect that these chains will be sufficiently long-lived to be used in quantum simulations of magnetic systems [E.E. Edwards, et al., Phys. Rev. B {\bf 82}, 060412(R) (2010)]. The trap is currently being fabricated at Georgia Tech using VLSI techniques.
\end{abstract}


\maketitle


\section{Introduction}

There have recently been a number of experiments and proposals that use linear chains of ions to simulate quantum systems or to perform quantum computation \cite{Kim09,Wel11,Edw10,Lin09}.  The ions are entangled through the modes of their collective motion, either by applying an optical spin-dependent force \cite{Kim09} or through an alternating magnetic field gradient \cite{Osp08,Wel11}.  A quantum simulation of the Ising model subject to a transverse magnetic field has been performed on three ions held in a linear radio frequency (rf) trap, coupled through the transverse motional modes \cite{Edw10}.  This three-ion system could be scaled to large numbers of ions by confining the ions in an anharmonic linear potential with nearly equal spacing between the ions \cite{Lin09}.  If the trap can be designed to hold thirty ions or more in a stable linear chain, it should be possible to study quantum systems that cannot be currently simulated using classical methods.  

Trapping long ion chains places a number of significant demands on the trap performance.  Generating an anharmonic potential for equally spaced ions requires a large number of segmented electrodes.  While it is possible to approximate this potential using as few as five electrode pairs \cite{Lin09}, maximizing inter-ion spacing uniformity could require as many as ten pairs of electrodes for longer chains.  While this can be realized using microfabricated surface-electrode ion traps, these traps have lower trapping depths and larger anharmonic terms in the radial trapping potential compared to traditional four-rod Paul traps. This can cause reduced ion lifetimes and increased mode frequency drift. Both of these issues become more important as the system is scaled up.

The trap design proposed in this paper features a novel monolithic multilevel symmetric structure that can be fabricated using processes similar to those used for microfabricated surface-electrode ion traps.  Electrostatic simulation studies of the proposed trap design indicate that it has an  order of magnitude deeper radial trapping potential and reduced radial anharmonicities in the rf confinement field as compared to that of surface-electrode traps.  We expect that the larger trap depth will increase ion chain lifetimes. The reduced radial anharmonicities will suppress mode frequency drifts associated with dielectric charging.  This trapping structure, therefore, combines the flexibility and scalability of silicon microfabrication technologies with the superior trapping characteristics of traditional four-rod Paul traps.

\begin{figure*}
\includegraphics{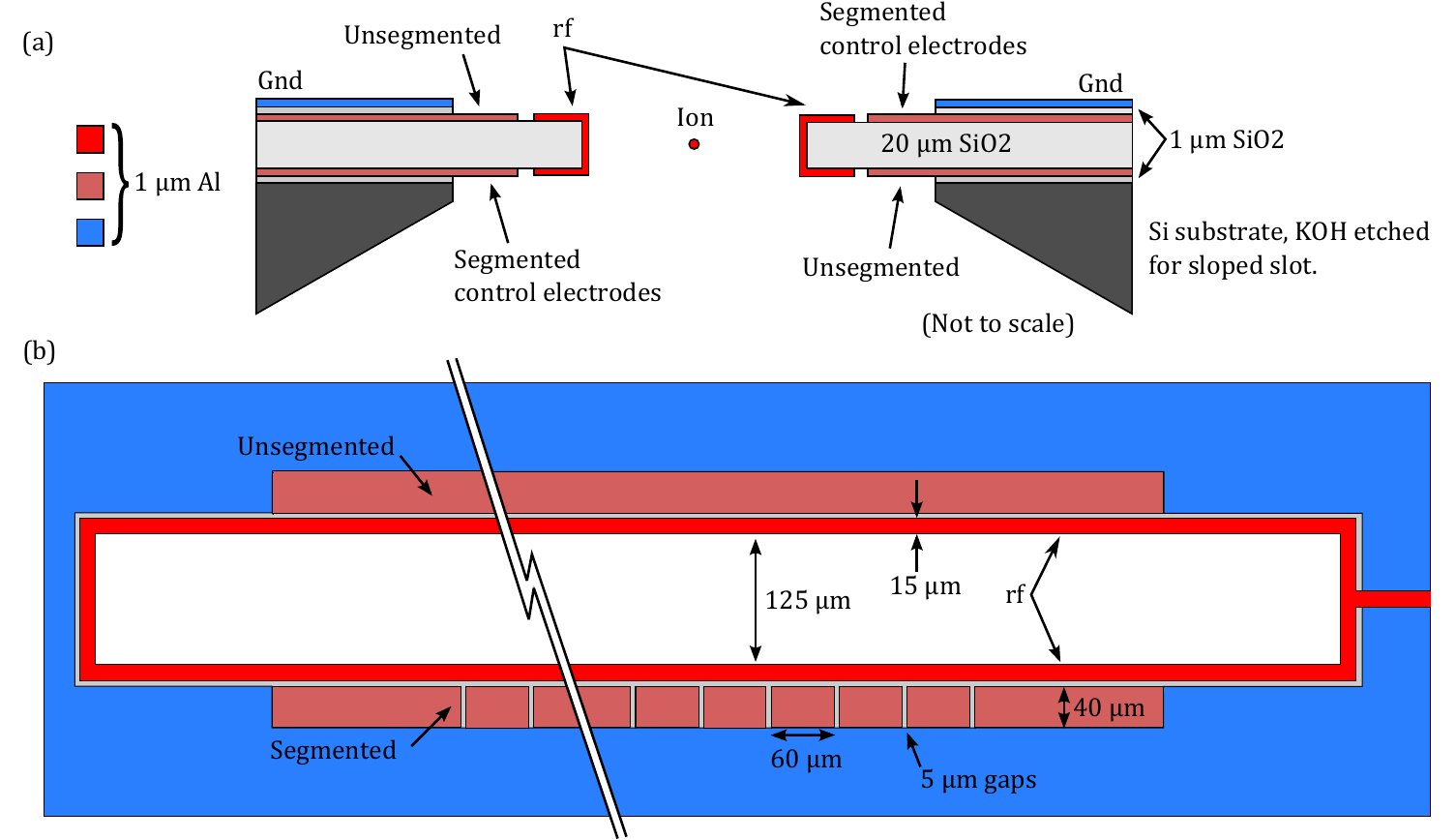}
\caption{Proposed symmetric trap described in this paper showing (a) the cross-section and (b) top view showing the layout of the dc control electrodes. The ion is trapped in the slot between the two rf rails. The Si substrate is a single wafer with a finite length slot etched through the wafer.}
\label{fig:trapdesign}
\end{figure*}

\section{Trap design}
A cross-section of the proposed design is shown schematically in Figure \ref{fig:trapdesign}a with the detailed dimensions for the trap electrodes shown in Figure \ref{fig:trapdesign}b.  The ion is trapped in the slot between the two rf electrodes. 

The beveled sides of the central slot provide through-chip optical access and can be formed using KOH etching of the Si substrate.  The dimensions of this slot were chosen to be as large as possible in order to reduce light scattering and consequent photoemission but not so large as to require excessively large rf and control voltages to maintain adequate trap depth.  For example, if the slot width were increased to 200~$\mu$m, the rf peak potential would have to be well above 300~V in order to maintain a 1~eV trap depth for \Yb ions at 40~MHz rf frequency.  In our experience, this potential would likely cause arcing from the rf electrode to nearby rf grounds. However, for a 125~$\mu$m slot the trap depth is about 1 eV for only 180~V applied rf.  The trap depth of 1 eV is an order of magnitude larger than the depths typically obtained for surface-electrode traps of similar size. Furthermore, the symmetric structure of the trap is expected to aid in stabilizing the radial mode frequencies by avoiding anharmonic terms in the rf pseudopotential that are present in surface-electrode traps.

The gap between electrodes is chosen to be 5~$\mu$m. This keeps the exposed oxide to a minimum while maintaining a large enough gap to prevent rf arcing. The metallic layers are all 1~$\mu$m thick Aluminum (Al).  The ground electrodes overlaying the dc control electrodes provide a capacitive rf short to ground, minimizing any rf pick up on the control electrodes, and protect the top surface from scratches.  A 1~$\mu$m thick dielectric (SiO$_2$) layer separates this ground from the control electrodes.

The dc control electrodes are segmented in 60~$\mu$m wide sections as shown in Figure \ref{fig:trapdesign}b.  The alternate diagonals are single long unsegmented electrodes that can be biased to aid in compensating the trap and controlling the angle of the principal axes for the ion motion. Although leaving the diagonal electrodes unsegmented somewhat restricts the control of the ion chain, it keeps the required number of electrode connections to below 50 and consequently simplifies the in-vacuum hardware needed to mount this trap. 

The setback of the dc electrodes from the rf electrode faces is 20~$\mu$m (15~$\mu$m rf width plus 5~$\mu$m gap).  This is a critical dimension for the design. It must be kept to a minimum; otherwise, the dc control voltages required to trap the chain exceed 20~V.  The thick 20~$\mu$m oxide layer, which forms the vertical face of the rf rail, provides a large trap depth for only 180~V applied rf peak (1~eV at 40~MHz rf frequency for \Yb) and also contributess to the mechanical stability of the structure.  Our studies show that reducing the oxide thickness to 10~$\mu$m will only result in a 20~\% reduction in trap depth but may physically weaken the structure.

\begin{figure*}
\includegraphics{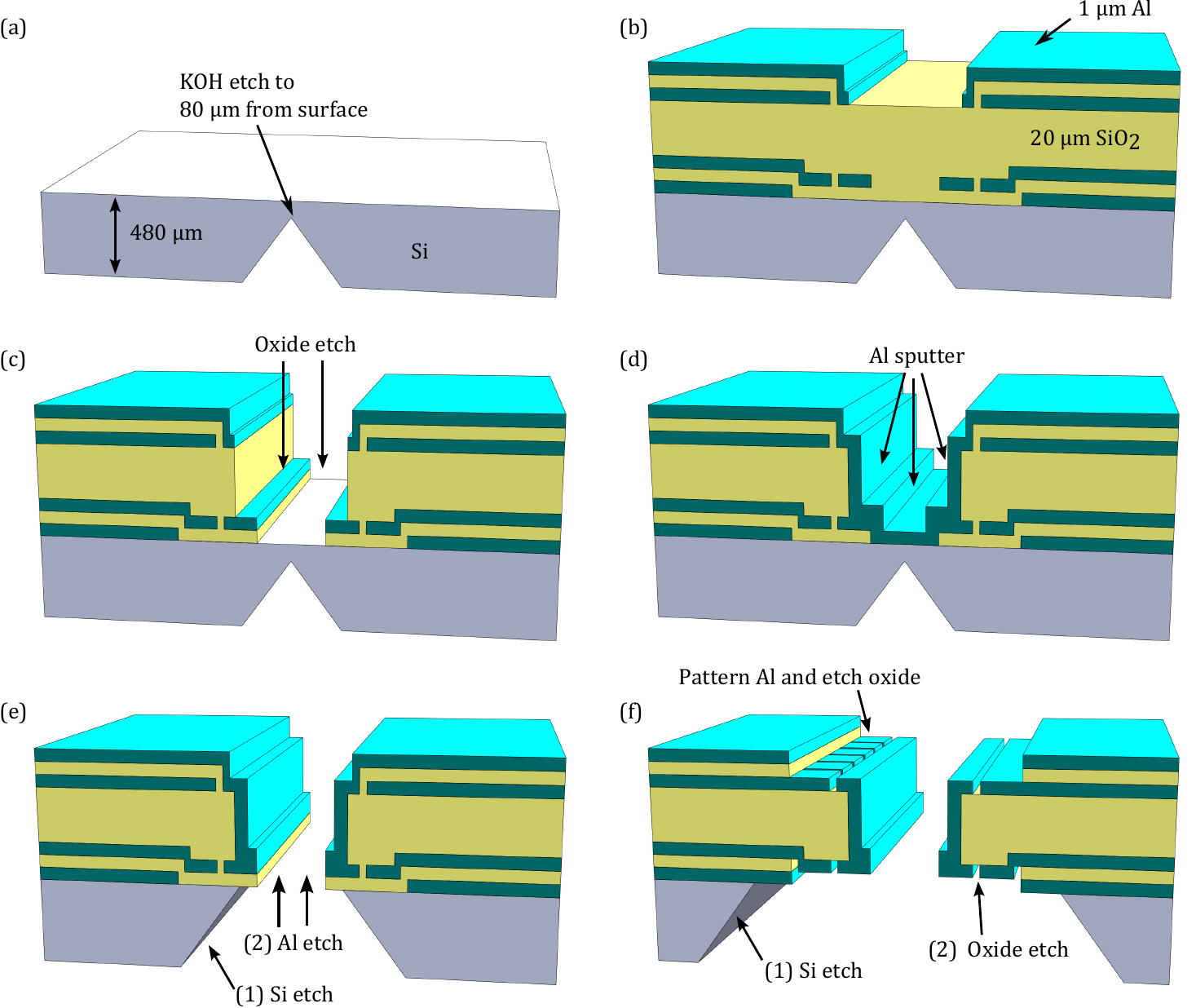}
\caption{Cross-section of the trap build-up as it goes through various fabrication steps.  See text for the descriptions of each step.}
\label{fig:trapfab}
\end{figure*}

The stability of the trapping frequencies is an important consideration for experiments using microfabricated ion traps. Charging of nearby exposed insulators can affect these frequencies.  The proposed design does not have a line-of-sight from the ion to any of the exposed oxide which is likely to reduce the influence of charges on these surfaces \cite{amini}.  Furthermore, compared to surface-electrode traps, the high symmetry in this design reduces anharmonicities in the radial pseudopotential and consequently reduces the dependence of the radial mode frequencies on stray fields created by the charging.

\section{Proposed fabrication process}

The trap design described in this paper is currently being micro-fabricated at the Nanotechnology Research Center (NRC) of Georgia Tech using standard Si based micro-fabrication technologies such as plasma enhanced chemical vapor deposition (PECVD), rf sputtering, dry plasma reactive ion etching (RIE), and contact lithography patterning.  The major steps are outlined below with corresponding diagrams in Figure \ref{fig:trapfab}a through f.

In (a), a silicon wafer is patterned on both sides with a SiN mask and a subsequent KOH etch forms a V shaped slot in the wafer that extends to within 80~$\mu$m of the top surface.  The SiN mask is removed and the alternating Al/oxide electrode structures are built up with standard microfabrication processes to form the cross-section shown in (b).  In (c) a portion of the metal above the apex of the KOH slot is patterned and a deep oxide etch removes the region that will become the central slot.  In (d), a crucial step in fabrication is accomplished by depositing a 1~$\mu$m thick Al metal layer on the sides and bottom of the etched central slot.  This deposition forms the inside vertical walls of the rf rails.  In (e), the KOH slot is enlarged and metal at the bottom of the central slot is etched away.  The bottommost oxide provides the mask for this metal etch.  Finally in (f), the KOH slot is further enlarged to expose the cantilevered trap arms and bottom control electrodes and a final oxide etch removes the exposed oxide.

\begin{figure*}
\includegraphics{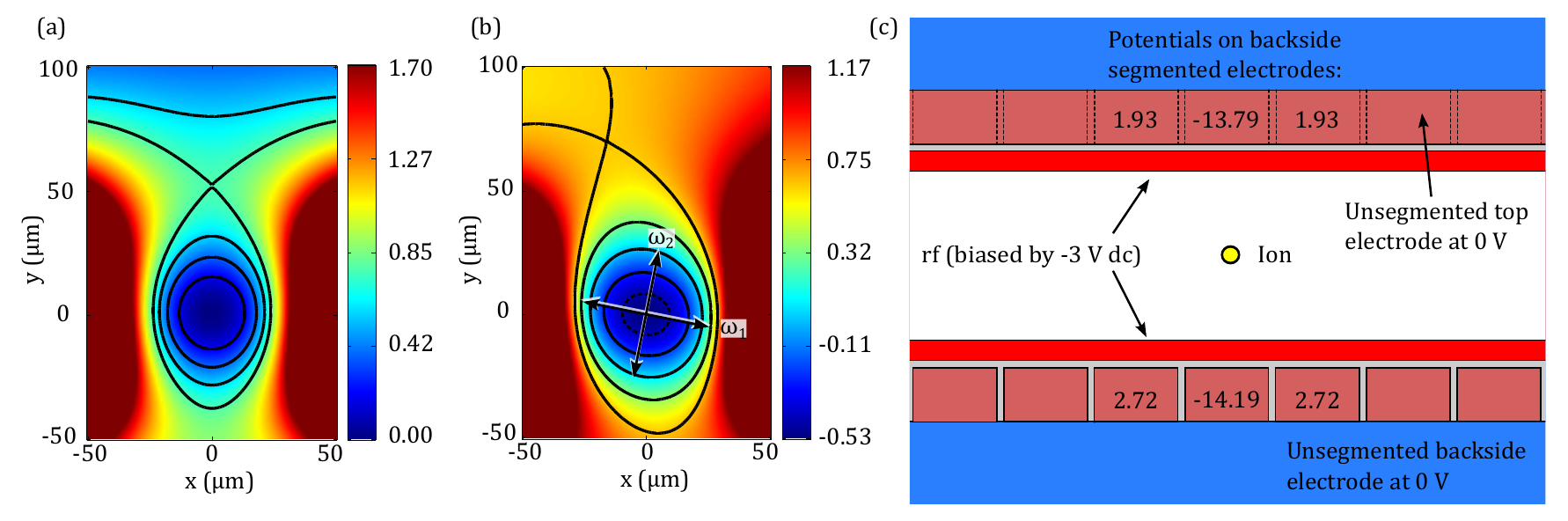}
\caption{(a) Radial trapping potential for a peak 180~V rf potential at 40~MHz rf frequency. The radial trap depth  is 0.76~eV, which corresponds the self-intersecting equipotential curve. The ion is trapped at $x=y=0$. (b) Total radial potential including the dc potentials shown in (c). The asymmetric dc potentials were designed to demonstrate rotation of the radial trap axis from the vertical, here shown by the two frequencies $\omega_1$ and  $\omega_2$ in (b).}
\label{fig:singleion}
\end{figure*}

\section{Trapping potentials for single ions}

Figure \ref{fig:singleion}a shows a radial cross-section of the pseudopotential in the proposed trap.  The static control potential configuration shown in Figure \ref{fig:singleion}c provides axial confinement for a single ion and lifts the degeneracy in the radial modes as shown in Figure \ref{fig:singleion}b.  The rf rails are biased by $-3$~V to increase the radial mode splitting without increasing the dc potentials with the rf peak potential of 180~V at 40~MHz.  The resulting radial mode frequencies for \Yb are 4.8~MHz and 5.8~MHz, at an angle of 11$^\circ$ to the vertical with a 1.0~MHz axial frequency. 

\begin{figure*}
\includegraphics{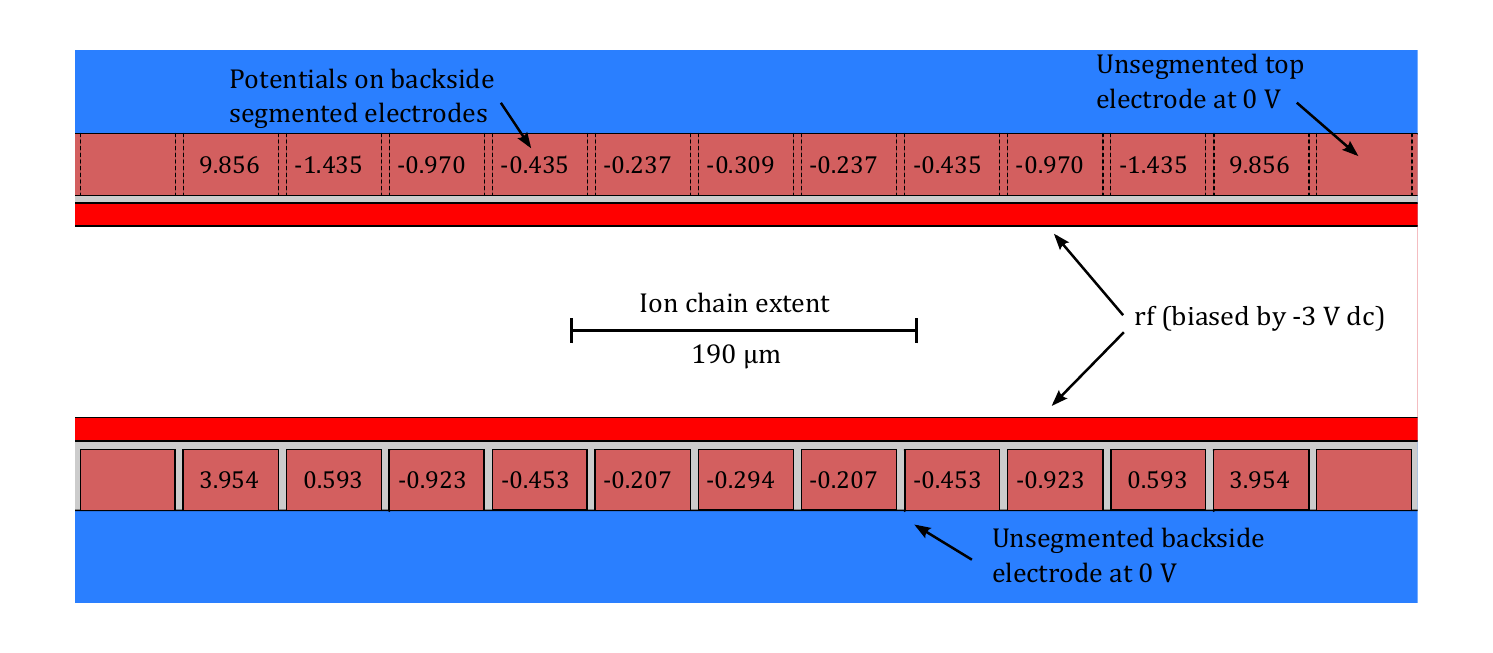}
\caption{Top schematic view of the symmetric trap is shown along with the segmented control electrode voltages optimized for equal spacing (10~$\mu$m) of a 20 ion chain. The rf electrodes are shown in green and the potentials are in volts.}
\label{fig:chainpot}
\end{figure*}

The slight top-bottom asymmetry in the trap radial potentials shown in Figure \ref{fig:singleion}a and \ref{fig:singleion}b is due to the silicon substrate. This results in a slightly weaker potential above the trap.  It is this direction that limits the trapping depth.  The radial well depth for the single trapped ion in Figure \ref{fig:singleion}b is 1.15~eV and exceeds the trap depth of the pseudopotential alone (0.76~eV) because the dc potentials provide additional confinement in the weakest direction. This trap depth is well above the typical depths for surface traps with comparable rf and dc potentials. 

\section{Ion chains}

The deep trapping depths in this design may be advantageous to holding long chains of ions with low loss rates.  We have calculated control potentials for holding ion chains with a uniform spacing.  The potentials are optimized to obtain (1) equal ion spacing, (2) constant principal axes along the chain,  (3) fixed strength of the radial dc quadrupole along the length of the chain, and (4) alignment of the dc field nulls with the rf null for each ion in the chain.  The radial dc quadrupole defines the radial mode frequency splitting and the rotation of the radial mode axes. 

\begin{figure}[H]
\includegraphics{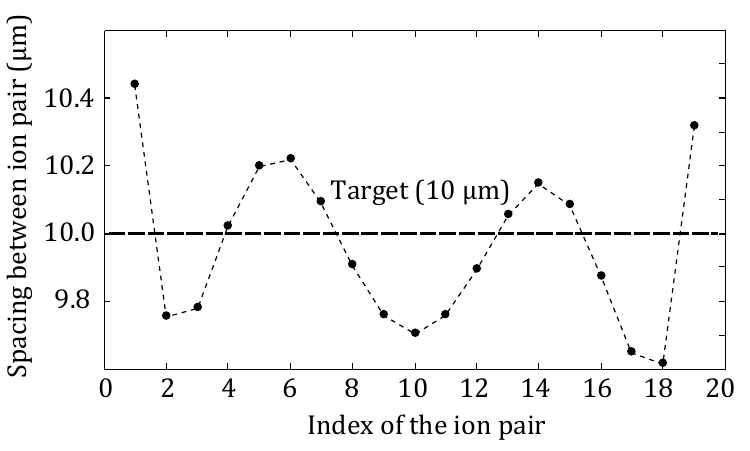}
\caption{Deviations from equal ion spacing across the 20 ion chain.}
\label{fig:spacing}
\end{figure}

\begin{figure}[H]
\includegraphics{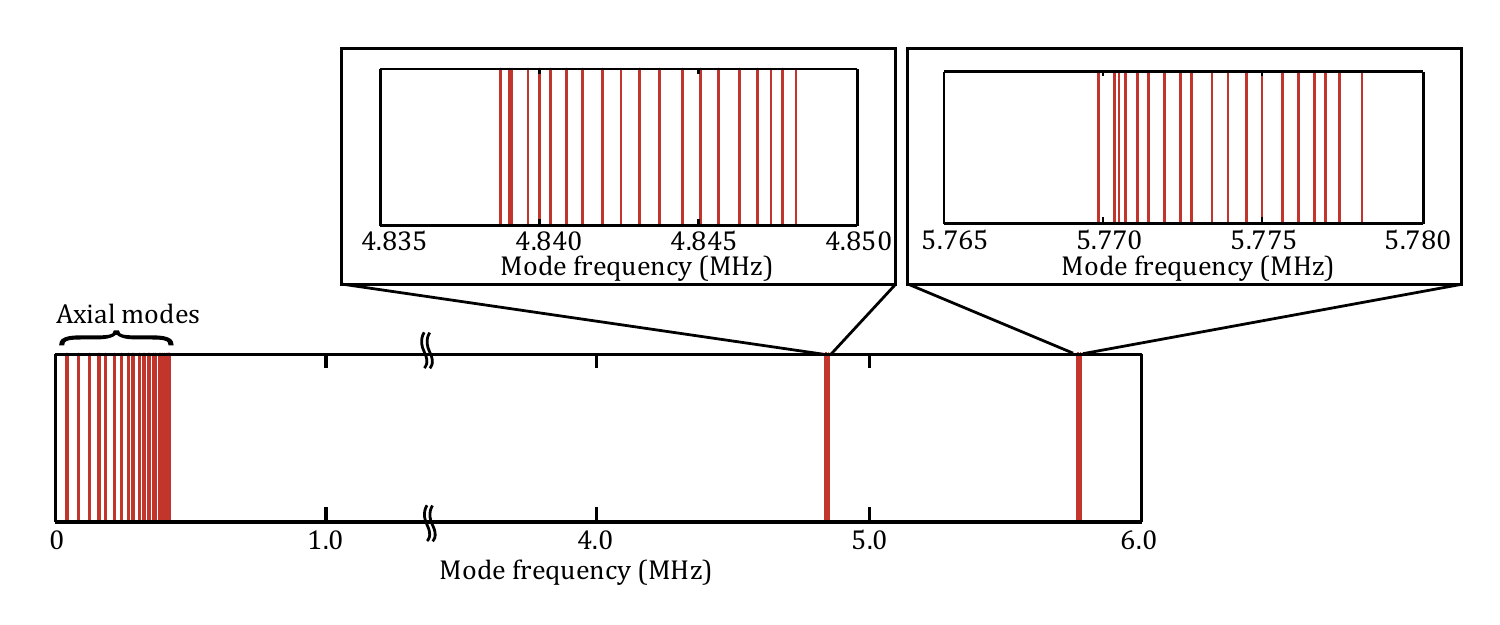}
\caption{Secular mode spectra for a 20 ion chain under the potentials for \Yb with a peak rf potential of 180~V, rf frequency of 40~MHz, and the control potentials shown in Figure \ref{fig:chainpot}.}
\label{fig:modes}
\end{figure}

An example set of potentials for a 20 ion chain meeting these conditions is shown in Figure \ref{fig:chainpot}.  The nominal ion spacing for this chain is 10~$\mu$m.  The calculated spacings vary from this goal by at most 0.5~$\mu$m with the maximum deviation at the end ions as shown in Figure \ref{fig:spacing}.  We have found that including three electrodes beyond each end of the chain, along with all the electrodes along the interior of the chain, is sufficient for the optimization procedure.  The axial and radial mode spectra for this 20 ion chain are shown in Figure \ref{fig:modes}.  Generating the large radial mode splitting was aided by biasing the rf rails by $-3$~V.  The angle of the transverse principle axes varies between 0.75~$^\circ$ at the chain ends to 0.61~$^\circ$ near the center of the chain for this simulation.  If required, a new set of control voltages could be generated that would set this angle to a larger value while keeping the variation small. 

We have studied the effect of dc electrode width on the ion chain control.  Improved control of ion spacing, principal axis orientation, and compensation are obtained by increasing the number of control electrodes and decreasing the electrode width.  From the results shown in Table \ref{tab:elecwidth}, it appears that decreasing the electrode width beyond 60~$\mu$m gives no additional benefit.  No significant gains are obtained for smaller widths.  However, we have not explored the full parameter space of electrode widths and more optimization may lead to smaller errors.

\section{Conclusion}

The novel trap design described here shows promise in combining the best elements of four-rod Paul traps, which have large trap depths and symmetric rf potentials, and microfabricated surface-electrode ion traps, which are amenable to scaling to large numbers of ions and can be fabricated with large numbers of electrodes for tailored trapping potentials.  The symmetric trap decribed in this paper has a radial well depth of 1.1~eV for \Yb  ion with an rf peak potential of 180~V at 40~MHz, which is an order of magnitude larger than that of surface-electrode traps of comparable size.  The deeper trapping depth will likely increase ion lifetimes for long chains, making it useful for many proposed quantum simulation and information processing experiments.  The proposed fabrication of this trap uses a similar set of tools and technology as Si based surface-electrode traps \cite{Lei09, amini, brownnutt} and is currently under fabrication at the Georgia Tech Research Institute. This work is supported by the DARPA OLE program under ARO award W911NF-07-1-0576. 

\onecolumngrid
\hspace{1in}

\twocolumngrid
\hspace{1in}

\pagebreak

\begin{table}[H]
\vspace{0.2in}
\begin{tabular}{|l|c|c|c|}
\multicolumn{4}{l}{(a) 20 ion chain with 10~$\mu$m spacing} \\
\hline
Electrode width & 500~$\mu$m	& 60~$\mu$m	& 30~$\mu$m \\
 (not including 5~$\mu$m gaps)	& & & \\ \hline
No. of active electrodes	& 5	& 5	& 7 \\ \hline
Max. spacing error	& 4.7~$\mu$m	& 0.75~$\mu$m	& 0.65~$\mu$m \\ 
\hline
\multicolumn{4}{l}{\vspace{0.2in}}\\
\multicolumn{4}{l} {(b) 50 ion chain with 10~$\mu$m spacing} \\
\hline
Electrode width & 500~$\mu$m	& 60~$\mu$m	& 30~$\mu$m \\
(not including 5~$\mu$m gaps)	& & & \\ \hline
No. of active electrodes &	5	& 7	& 10 \\ \hline
Max. spacing error & 12~$\mu$m	& 1.5~$\mu$m	& 1.6~$\mu$m \\
\hline
\end{tabular}
\caption{Spacing errors as a function of electrode width for (a) 20 and (b) 50 ion chains. The target ion spacing is 10~$\mu$m for both cases. }
\label{tab:elecwidth}
\end{table}

\onecolumngrid

\pagebreak
\twocolumngrid

\bibliography{SymmetricTrapBib_v1}

\end{document}